%% file: swap_hhl.tex
\documentclass[conference]{IEEEtran}
\IEEEoverridecommandlockouts
\usepackage[numbers,sort&compress]{natbib}
\usepackage{amsmath,amssymb,amsfonts}
\usepackage{graphicx}
\usepackage{braket}
\usepackage{subcaption}
\usepackage[usenames,dvipsnames,table,xcdraw]{xcolor}
\usepackage{tabularx}
\usepackage[hidelinks]{hyperref}
\usepackage{tikz}
  \usetikzlibrary{quantikz}
\usepackage{fontawesome}

\begin{document}

\title{Accelerating the training of single-layer binary neural networks using the HHL quantum algorithm}

\author{\IEEEauthorblockN{Sonia Lopez Alarcon, Cory Merkel, Martin Hoffnagle, Sabrina Ly}
\IEEEauthorblockA{\textit{Department of Computer Engineering} \\
  \textit{Rochester Institute of Technology}\\
  Rochester NY, USA \\
  \{slaeec, cemeec, mah5414, sjl2178\}\faAt{}rit.edu}
\and
\IEEEauthorblockN{Alejandro Pozas-Kerstjens}
\IEEEauthorblockA{\textit{Institute of Mathematical Sciences}\\
  \textit{(CSIC-UCM-UC3M-UAM)}\\
  Madrid, Spain \\
  physics\faAt{}alexpozas.com}

\thanks{This work is partially supported by the Spanish Ministry of Science and Innovation MCIN/AEI/ 10.13039/501100011033 (``Severo Ochoa Programme for Centres of Excellence in R\&D'' CEX2019-000904-S and grant PID2020-113523GB-I00), the Spanish Ministry of Economic Affairs and Digital Transformation (project QUANTUM ENIA, as part of the Recovery, Transformation and Resilience Plan, funded by EU program NextGenerationEU), Comunidad de Madrid (QUITEMAD-CM P2018/TCS-4342), and the CSIC Quantum Technologies Platform PTI-001.}
}

\maketitle

\begin{abstract}
Binary Neural Networks are a promising technique for implementing efficient deep models with reduced storage and computational requirements. The training of these is however, still a compute-intensive problem that grows drastically with the layer size and data input. At the core of this calculation is the linear regression problem. The Harrow-Hassidim-Lloyd (HHL) quantum algorithm has gained relevance thanks to its promise of providing a quantum state containing the solution of a linear system of equations. The solution is encoded in superposition at the output of a quantum circuit. Although this seems to provide the answer to the linear regression problem for the training neural networks, it also comes with multiple, difficult-to-avoid hurdles. This paper shows, however, that useful information can be extracted from the quantum-mechanical implementation of HHL, and used to reduce the complexity of finding the solution on the classical side.
\end{abstract}


\input{01_introduction}
\input{02_background}
\input{03_1_AcceleratedBNN}
\input{03_2_QbinaryNN}
\input{03_2_Probability}
\input{04_results}
\input{05_conclusion}

\bibliographystyle{IEEEtran}
\bibliography{biblio}

\end{document}

%% file: 01_introduction.tex
\section{Introduction and Motivation}
Machine Learning and in particular deep learning and neural networks, are almost ubiquitous in today's world, from finances to healthcare. The growth of data and the size of the models, however, place increasing pressure on the computational resources that support them, and they constantly fall short of growing expectations and demands.

Although still in very early stages, Quantum Computing holds some promises of mitigating this issue, even if as a hybrid model that will partially solve the problem when combined with classical approaches. Current trends in this direction include the Quantum Approximate Optimization Algorithm, QAOA \cite{qaoa}. An optimization problem is at the core of any neural network training process, with the goal of finding the weights to be assigned to the nodes of the network. QAOA attempts to solve this optimization problem by combining implementing a paramatrizable quantum circuit for which parameters are adjusted to find the solution within a predefined cost function. It is possible to implement this approach within the Noisy, Intermediate-Scale Quantum computing era~\cite{nisq}, which unfortunately only provides approximate solutions for, at the moment, very small models. However, the fact that current quantum computing devices are physical systems subject to noise can also be exploited for aiding the training of classical machine learning models~\cite{Kehoe2021}.

On the other hand, the optimization problem associated with the training of single-layer neural networks can be approximated as a linear regression problem. The HHL quantum algorithm \cite{HHL,bayesiandl} was proposed as a potential solution linear systems of equations. However, plainly using HHL to solve the training of NN is unfeasible due to a number of reasons \cite{readthefineprint}, even if the technology progresses to provide sufficient resources and low noise levels.

One of the hurdles is that HHL provides, at the output of the quantum circuit, a quantum state that encodes the solution of the linear system of equations in superposition. Extracting the value of each of the components of the solution vector requires in the best case scenario ---i.e., the solution being in the $+\mathbb{R}$ field--- to perform many measurements in the computational basis, which allegedly ruins the potential quantum advantage. The number of measurements required (computed as the number of runs of the protocol times the number of qubits measured at each run) to accurately extract the amplitudes of a superposition state with coefficients in $+\mathbb{R}$ is generally assumed to be in the order of $n$, where $n$ is the dimension of quantum system, or in the particular case of neural network training, the number of weights. This, however, depends on the accuracy that is expected of the solution, and the shape of the outcome probability distribution. If the solution vector is not in the $+\mathbb{R}$ field, the situation is even worse, requiring quantum tomography to extract the amplitudes of the quantum states in superposition. This problem is discussed in more depth in Section \ref{sec:ProbabilityDistribution}.

It is possible, however, to apply operations to the solution in superposition. One of such operations is the SWAP test \cite{swaptest}, which can be used to extract the distance between two quantum states, encoding it in the amplitudes of a single qubit as will be shown in Section \ref{sec:AccelTraining}.  In a similar context \cite{bayesiandl}, the SWAP test has been used in the past to verify the correctness of proof of concept implementations of HHL. When the implementation is correct, the distance with the foreknown solution state is equal to zero. This is, however, not useful unless one is able to calculate the solution ---classically--- beforehand, and this solution is implementable as a quantum superposition.

This paper proposes that, instead of being used to compare the circuit's preparation with the solution (which implies knowing the solution in advance), \emph{the SWAP test be performed against several, fixed, ``reference'' states, and the information obtained is used to reduce the computational complexity of the classical search for the optimal set of weights in the training process of  Binary neural networks (BNN)}.

BNNs are promising as possible computationally efficient implementation to neural networks. BNNs are the  specific realization of quantized NN in which the weights will be limited to one of two values 1 or -1, or for the implementation targeted in this paper, 0 or 1. BNNs are also promising as a possible candidate to the application of HHL in combination with the SWAP test to extract information for two reasons: (1) the solution is promised to be in the $+\mathbb{R}$ field, and (2) the vector to compare against is easy to implement as a quantum superposition state.

\begin{figure}
    \centering
    \includegraphics[width=2.0 in]{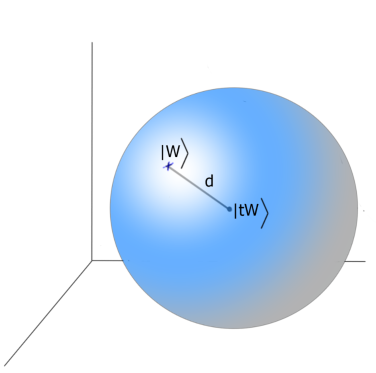}
    \caption{The proposed quantum implementation allows to use the SWAP test to calculate the distance between the weights state ---state containing the solution to the optimal weights vector--- $\ket{W}$ and some other selected test state $\ket{tW}$. The search space is then reduced from the hyperspace of all possible solutions to the surface of the hyper-sphere with center in $\ket{tW}$ and radius $d$.}
    \label{fig:hypersphere}
\end{figure}

Figure \ref{fig:hypersphere} illustrates this approach. As this paper will show, given a ``reference'' test, the SWAP test can provide the distance from the solution to the this state, and hence, reduce the classical search of the solution to the hypersphere of distance $d$ from the reference state.

%% file: 02_background.tex
\section{Background and Related Work}\label{sec:background}
\subsection{Binary Neural Networks}\label{sec:bnn}

Machine learning continues to make rapid advances in its ability to accurately model complex relationships in high-dimensional data. A prime example is the application of deep artificial neural networks (ANNs) for classification of objects in images.  From automatic mail sorting to self-driving cars, this capability is a key enabler to automate some of the most critical aspects of human decision making.  However, the size and complexity of ANNs performing these operations appears to grow exponentially with linear gains in their performance \cite{hendy2022review}.  This has led to top-performing models with billions of parameters that require massive computational resources to train and evaluate.  Consequently, there is a push to reduce the size of ANN models to enable their deployment on resource-constrained hardware such as edge devices.  While several methods exist, one of the most popular approaches is to reduce the precision of ANN computations and parameters.  In the extreme case of binary neural networks (BNN), the ANN weights and activation functions are reduced from 32- or 64-bit floating point values to 1-bit representations (e.g. -1 and 1, or 0 and 1).

\begin{figure}
    \centering
    \includegraphics[scale=0.5]{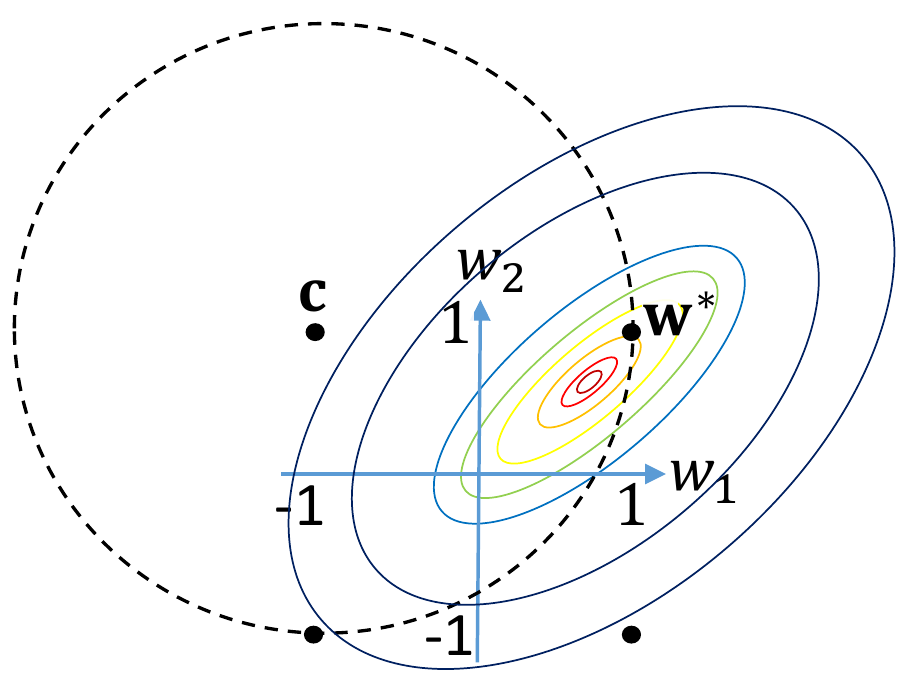}
    \caption{Gradient descent constrained to the surface of a hypersphere.  Ovals indicate lines of equal loss value, with hotter colors corresponding to lower loss.}
    \label{fig:classicsearchwithhypersphere}
\end{figure}

BNNs are usually trained using quantization-aware methods, where a full-precision model is quantized during the forward pass and then updated using backpropagation with approximations for the gradient of the quantization operator.  Given an ANN model $\Pi$ and a loss function $\mathcal{L}$, a weight connecting neuron $i$ in layer $l-1$ and neuron $j$ in layer $l$ gets updated by gradient descent:
\begin{equation}
\Delta w_{i,j}^{l}=-\alpha\frac{\partial\mathcal{L}\left\{q(\Pi)\right\}}{\partial w_{i,j}^{l}}
\end{equation}
where $q$ is the quantization function and $\alpha$ is the learning rate. The quantization function can take several forms. The simplest one is rounding the corresponding value (weights or activations) to the nearest valid quantization level.  During training, the forward and backpropagation steps require matrix-vector multiplications at each layer.  Here, we assume that the computation of the binary activation function, which requires only a scalar comparison operation, is much less costly than the matrix-vector product. Thus, the time complexity approximately scales as $\mathcal{O}\left(\sum\limits_{l=1}^{L}N_{l}N_{l-1}\right)$, where $L$ is the total number of layers and $N_{l}$ is the number of neurons in layer $l$.  This can be bounded from above as $\mathcal{O}\left(kLN_{max}^{2}\right)$, where $N_{max}$ is the maximum layer size and $k$ is the number of training iterations.  Generally the complexity of $k$ is $\mathcal{O}(1/\epsilon)$, where $\epsilon$ is the loss level.  If information about the solution is known a priori, then the gradient descent can be constrained, possibly leading to a smaller $k$.  For example, if the Euclidean distance $d$ to the solution $\mathbf{w}^{*}$ from a test point $\mathbf{c}$ is known, then the search space becomes restricted to the hypersphere centered at $\mathbf{c}$.  This is demonstrated in Figure \ref{fig:classicsearchwithhypersphere}.  Note that the intersection of the hypersphere with the vertices of the hypercube form a hyperplane of form
\begin{equation}
\label{eq:plane}
\mathbf{c}\cdot\mathbf{w}=N-\frac{d^{2}}{2}
\footnote{Respecting the ML and QC notations, $\mathbf{w}$ and $\mathbf{c}$  correspond to $\ket{W}$ and $\ket{tW}$ respectively in Figure \ref{fig:hypersphere} and other sections of this paper.}
\end{equation}

Later, we will show that restricting the search space of an $N$-dimensional linear least squares problem to an $N-1$-dimensional hyperplane, and within the boundaries of the hypercube defined by the possible solutions of the weights vector, can reduce the average number of iterations required to find the solution.

\subsection{HHL algorithm for solving linear systems of equations}\label{sec:HHL}
In 2009, Harrow, Hassidim and Lloyd published a paper describing a quantum algorithm for solving linear systems of equations \cite{HHL}. The problem can be defined as, given a matrix $A$ and a vector $\vec{b}$, find a vector $\vec{x}$ such that $A\vec{x} = \vec{b}$. In quantum notation, this is linear system of equations is expressed as $$A\ket{x}=\ket{b},$$
\noindent where $A$ is a Hermitian operator --- a workaround exists when $A$ is not Hermitian--- and $\vec{b}$ has to be encoded in a quantum state $\ket{b}$ and, hence, it has to be normalized. The solution to this linear system of equations is, therefore, expressed as $$\ket{x}=A^{-1}\ket{b}.$$ The algorithm assumes the case being considered is one in which the solution $\ket{x}$ does not need to be fully known, but rather an approximation of the expectation value of some operator associated with the solution vector.
 \par
The problem boils down to finding $A^{-1}$, like in any other algorithm to solve linear systems of equations. Classical algorithms usually have a computational complexity $O(N^3)$, where $N$ is the size of $A$. If the operator were expressed as a diagonal matrix, the calculation of $A^{-1}$ is almost immediate, creating a diagonal matrix in which the eigenvalues $\lambda_i$ in the diagonal are replaced by their corresponding inverted values, $\lambda_i^{-1}$. The computationally intensive portion of this solution in the classical context is precisely to calculate the eigenvalues of $A$. This, however, can be solved rather easily in a quantum-mechanical manner through the use of the quantum phase estimation in which the eigenvalues are calculated and expressed as a phase.

The HHL algorithm can potentially estimate the function of the solution vector in running time complexity
$O(\log(N)s^{2}k^{2}/\epsilon)$ given that the matrix $A$ is $s$-sparse and well-conditioned, where $k$ denotes the condition number of
the system and $\epsilon$ the accuracy of the approximation~\cite{Qiskit}.

\begin{figure}[ht!]
 \centering
  \includegraphics[width=0.5\textwidth]{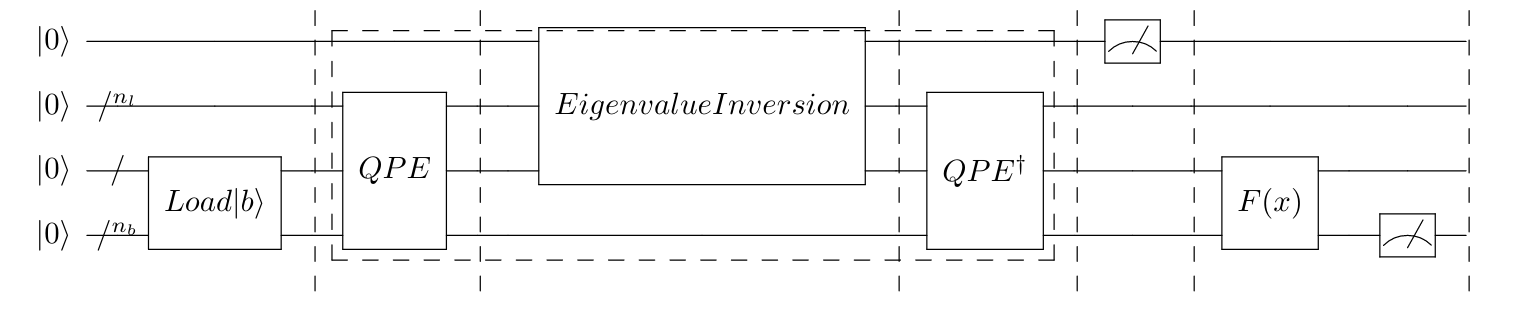}
  \caption{Generic HHL Algorithm with 5 stages: state $|b\rangle$ preparation,
  Quantum Phase Estimation, Eigenvalue
  Inversion, inverse QPE, measurement of the flag qubit, and application of
  some function $F(x)$. Note that only in the case of the measurement in the first register producing the outcome $1$ the algorithm succeeds~\cite{HHL}.}
\label{qiskit_HHL}
\end{figure}

The circuit implementation can be broken down into six different circuit stages: loading the $|b\rangle$
input, Quantum Phase Estimation, Eigenvalue Inversion, inverse Quantum Phase Estimation, ancilla
qubit measurement, and application of a function $F(x)$ followed by measurement as shown in Figure \ref{qiskit_HHL}. $F(x)$ can be any linear equation in which some quantum mechanical operator is applied in order to obtain an estimate of the expectation value.
\par

\subsection{The SWAP test}\label{sec:swap}
\begin{figure}
  \centering
  \begin{quantikz}
    \lstick{$\ket{0}$} & \gate{H} & \ctrl{2} & \gate{H} & \meter{} \\
    \lstick{$\ket{A}$} & & \swap{1} & \qw & \qw \\
    \lstick{$\ket{B}$} & & \targX{} & \qw & \qw
  \end{quantikz}
  \caption{The standard SWAP test. The resulting state of the circuit is (without normalizations) $\ket{0}(\ket{AB}+\ket{BA}) + \ket{1}(\ket{AB}-\ket{BA})$, and thus the probability of obtaining outcome $\ket{1}$ on the measurement is $\frac{1}{4}(\bra{AB}-\bra{BA})(\ket{AB}-\ket{BA})=\frac{1}{2}-\frac{1}{2}|\braket{A|B})|^2$, where if $\ket{A}=\sum_{i}a_i\ket{i}$ and analogously for $\ket{B}$, then $\braket{A|B}=\sum_i a_i^* b_i$. This allows computing the overlap of $\ket{A}$ and $\ket{B}$ with additive error $\epsilon$.}
  \label{fig:swap}
\end{figure}
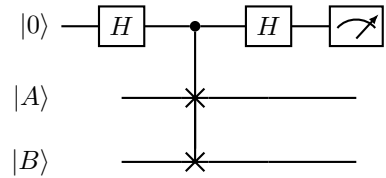

The \textit{SWAP test} routine is a quantum algorithm that expresses the scalar product of
two input states ~\cite{swaptest,Fastovets2019}. Two input states $|A\rangle$ and $|B\rangle$ are compared using controlled-SWAP gates and a control bit. The circuit for a two state comparison is shown in Figure \ref{fig:swap}.

The ancillary or control qubit is independent of the dimension of the state register and
the application of the first Hadamard gate moves the qubit from $|0\rangle$ to the
superposition state $|+\rangle$ \cite{multipleswaptest}. The CSWAP gates will then
exchange the pair of input states $|A\rangle$ and $|B\rangle$ if the
control qubit is in state $|1\rangle$. The second Hadamard gate will transform the
input pair into a superposition of symmetric and anti-symmetric state coupled with the
state of the ancillary qubit~\cite{multipleswaptest}.
The probabilities of measuring the the ancillary qubit as $\ket{0}$ and  $\ket{1}$ relate to the inner
product of the two states:

\begin{equation}
  \begin{aligned}
      p_{0} &= \frac{1}{2} + \frac{|\braket{A|B}|^{2}}{2},
      ~~p_{1} = \frac{1}{2} - \frac{|\braket{A|B}|^{2}}{2}
  \end{aligned}
  \label{eq:swapprobs}
\end{equation}

\noindent where $\braket{A|B}$ represents the inner product of the two quantum states $\ket{A}$ and $\ket{B}$.

The overlap provides information about the relationship of the two states. If the outcome probability in the measured qubit is a uniform distribution, the states $\ket{A}$ and $\ket{B}$ are orthogonal. Conversely, if the output $\ket{0}$ is measured with 100\% probability, the two states are equal. In addition,
the SWAP test can be used to calculate the Euclidean distance between classical  vectors encoded as quantum states.
\cite{Fastovets2019}.

\subsection{Related work}
The computational requirements for training ANNs and the large scale of the datasets that are typically employed in training have motivated many quests for finding faster and cheaper algorithms.
These include, as discussed in Section~\ref{sec:bnn}, the reduction of the precision in the ANNs' parameters, but also novel ways of computing by exploiting the physical principles of quantum mechanics.

A popular approach to exploit quantum computing in the training of ANNs is the formulation of training of Quadratic Unconstrained Binary Optimization (QUBO) problems~\cite{date2021}.
This choice is motivated by the existence of quantum annealers \cite{dwave}, which are physical systems whose ground state encodes the solution of such problems.
Therefore, time evolution under suitable conditions naturally drives the system to the solution.
In contrast to superconducting quantum circuits, quantum annealers already feature thousands of qubits, and it is generally accepted that the process uses quantum tunneling to overcome high energy barriers and increment the probability of obtaining the ground state~\cite{boixo2016tunneling}.
Recently, the training of BNNs has been formulated as a QUBO problem, enabling its solution in quantum annealers~\cite{sasdelli2021}, and showing promising results in state-of-the-art hardware.
Moreover, the appearance of Ising machines~\cite{isingmachine2016,isingmachine2019}, which encode the same type of problems in classical physical systems, implies that quantum effects are not a necessity for obtaining good solutions to QUBOs.

Another quantum-based technique for training ANNs is found in Ref.~\cite{bayesiandl}.
There, the authors exploit the connection between the training of deep neural networks and of Gaussian processes \cite{nnandgp1,nnandgp2} in order to provide a quantum algorithm for training deep ANNs.
In order to compute the mean and the variance of the corresponding Gaussian process predictor, Ref.~\cite{bayesiandl} leverages the HHL algorithm and quantum inner product subroutines.
By quantizing the necessary operations, the authors obtain an end-to-end algorithm that does not need of extracting individual entries of the HHL solution vector at any stage (for the case of neural networks with ReLU activation functions).
Moreover, Ref.~\cite{bayesiandl} provides small-scale demonstrations of the HHL subroutine, benchmarking it in classical simulators of quantum circuits and in the cloud-based quantum computers available at the time.
In order to evaluate the accuracy of the subroutine when executed in real quantum processing units, the authors perform a SWAP test with the target state, known in advance, and prepared in an auxiliary register.
This allowed to extract the information about the quality of the preparation by measuring just one qubit of the circuit in the standard basis, instead of having to perform tomography on the full circuit.
With a slightly different goal (namely estimating the distance from the prepared state to a number of fixed, target states), this advantage is the one we exploit in the remainder of this article.

%% file: 03_1_AcceleratedBNN.tex
\section{Accelerated Training}\label{sec:AccelTraining}
The proposed solution to the problem of training of NN is a hybrid Quantum-Classical solution as follows:
\begin{itemize}
    \item Given the data for training a NN, the linear regression problem is expressed as system of linear equations.  The optimal weights for this NN can be obtained by HHL, which encodes them in the amplitudes of a quantum state $\ket{W}$. In particular, in the case of a BNN, the fact that $\ket{i}$ is in the final state implies that the neuron $i$ is active (i.e., that the weight $w_i$ is $1$).
    \item In order to access which amplitudes are nonzero, quantum tomography in the computational basis is necessary to recover the full state, but this is computationally expensive (see Section \ref{sec:ProbabilityDistribution}).
    \item However, if an additional state (test state $\ket{tW}$) is wisely selected and encoded to perform a SWAP test against, useful information can be extracted. Measuring the ancillary qubit at the output of this SWAP test is not computationally expensive (see Section \ref{sec:ProbabilityDistribution}).
    \item The information is then passed to the classical side of the algorithm to solve the final problem, which results in an optimization problem in a much reduced space.
\end{itemize}

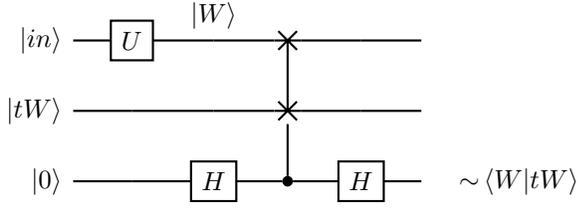
\begin{figure}
  \centering
  \begin{quantikz}
    & & \ket{W} & & & & \\ [-0.6cm]
    \lstick{$\ket{in}$} & \gate{U} & \qw & \targX{1} & \qw & \qw & \\
    \lstick{$\ket{tW}$} & \qw & \qw & \swap{-1} & \qw & \qw &  \\
    \lstick{$\ket{0}$}  & \qw & \gate{H} & \ctrl{-1} & \gate{H} & \qw & {\displaystyle \sim } \braket{W|tW}\\
  \end{quantikz}
  \caption{High-level schematic of the proposed quantum circuit, where $U$ represents the implementation of the HHL circuit solving the linear regression of a given binary neural network. $\ket{W}$ is the quantum superposition that contains all the weights of such solution. $\ket{tW}$ is the quantum state chosen as test state. The result of the SWAP test is related to the overlap $\braket{W|tW}$.}
  \label{fig:swaptest}
\end{figure}

This approach of extracting the solution state $\ket{W}$ out of the HHL circuit ($U$) to then perform a SWAP test against a test state $\ket{tW}$ is summarized in Figure \ref{fig:swaptest}. The information that can be extracted out of the output of the SWAP test can be easily related to the overlap of the two quantum states via Eq.~\eqref{eq:swapprobs}.

Although the approach described above can be applied to solve NN with any weight values, BNNs are specially suitable for this approach for two reasons:
 \begin{itemize}
     \item The weights taking 1 or 0 values implies that the superposed state solution out of the HHL implementation --- encoding in its amplitudes approximately the normalized weights vector--- is guaranteed to be within the $+\mathbb{R}$ field.
     \item The test state can be easily encoded, simply using Hadamard gates.
 \end{itemize}

%% file: 03_2_QbinaryNN.tex
\subsection{Relevant information for acceleration}\label{QBNN}
Given the approach described above, this section discusses  the kind of information that can be extracted out of the quantum portion of the hybrid solution.

\begin{table}[h]
  \centering
  \caption{Variable definition}
  \begin{tabular}{||c c||}
   \hline
   Variable & Definition\\ [0.5ex]
   \hline\hline
    $\ket{W}$ & quantum superposition of weights vector \\
   \hline
   $\ket{tW}$ & quantum superposition of test weights vector \\
   \hline
    $\ket{w_i}$ & basis representing each of the NN weights \\
     \hline
    $\omega_i$ & amplitude of each $\ket{w_i}$\\
   \hline
   $W$ &  dimension of  $\ket{W}$ and $\ket{tW}$\\ & (total number of weights) \\
   \hline
  \end{tabular}
\end{table}

The problem of training BNNs is the problem of finding the weights $W$ that minimize a certain cost function with the restriction that those weights can only take two values, 0 or 1.
The vector of weights will have $N$ values equal to 1 and $W-N$ values equal to 0.

In the quantum representation, the weights vector is
\begin{equation}
   \ket{W}=\sum_i^W\omega_i\ket{w_i},
\end{equation}

\noindent where the $p(\ket{w_i})=|{\omega_i}|^2$  probabilities of measuring he state $\ket{w_i}$  is equal to the value of the weight $w_i$. This is accurate when the probabilities are measured in the computational basis, and the weights are real and positive. This is the case for a BNN. Furthermore, in this case in which the probabilities are equal to either 1 or 0, and given that the quantum state is normalized, the weight vector can be expressed with a uniform distribution where $\omega_i=\frac{1}{\sqrt{N}} $ if  $w_i=1$, and $\omega_i=0$ if  $w_i=0$.

Because of this, we will choose to use test weight vectors, $\ket{tW}$, that have the form of uniform superpositions as well.
One of the reasons why BNNs can benefit form this approach is because of the low computational complexity of preparing states in uniform superposition.

\subsection{The SWAP test and the Euclidean distance}
Let us assume a quantum \emph{test state} $\ket{tW}$ representing a test vector with $N$ weights $w_i$ with value equal to 1.
The distance of this state to the target state, $\ket{W}$, is given after applying the SWAP test as described in Figure \ref{fig:swaptest}. The resulting state in the ancillary qubit is given by

\begin{equation}
   \ket{d}=\frac{1}{2}\ket{0}(\ket{W,tW}+\ket{tW,W}) +\\ \frac{1}{2}\ket{1}(\ket{W,tW}-\ket{tW,W})
\end{equation}

 According to this, the probability of finding $\ket{d}= \ket{0}$ is given by

 \begin{equation}
      P_{\ket{d}=\ket{0}}= \frac{1}{2}+\frac{1}{2}|\braket{W|tW}|^2
  \end{equation}

It is known that the Euclidean distance of two vectors $\vec{a}$ and $\vec{b}$ is  expressed  as
  \begin{equation}
     Ed^2=|\vec{a}|^2 + |\vec{b}|^2 -2\vec{a}\vec{b}
  \end{equation}

  where term $\vec{a}\vec{b}$ is the inner product of the two vectors, and represents the overlap between the two. The higher this overlap, the smaller the distance between the two.

  The information in terms of quantum states is contained in the $\frac{1}{2}|\braket{W|tW}|^2$ term of the $\ket{d}$ state. Considering that both $\ket{W}$ and $\ket{tW}$ are normalized, the Euclidean distance of the normalized vectors $Ed^2=2-2\braket{W|tW}$. When $\braket{W|tW}=0$ the states are orthogonal, while $\braket{W|tW}=1$ means that the normalized vectors are equal.

 Hence, the first piece of information that can be extracted is that the  solution weights vector to the BNN training problem and the test vector meet the relationship of a hypersphere with center in $\ket{tW}$ and radius $Ed^2$:

\begin{equation}
\sum_i(\omega_i-t\omega_i)^2=Ed^2,
\end{equation}
\begin{equation}
\sum_i(\omega_i-t\omega_i)^2=2-2\braket{W|tW}.
\end{equation}

Adding $s$ SWAP tests will reduce the solution to the intersection of $s$ hyperspheres of the form
\begin{equation}
\sum_i(\omega_i-t\omega_{i,j})^2=2-2\braket{W|tW_j}=Ed_j^2,
\end{equation}
where $j$ indicates the different test weights vectors from 1 to s.

\subsection{Multiple SWAP tests}

\begin{figure}
  \centering
  \subfloat[\label{fig:multiple1}]{
      \begin{quantikz}
        \lstick{$\ket{0}$} & \gate{H} & \ctrl{3} & \qw & \gate{H} & \meter{} \\
        \lstick{$\ket{0}$} & \gate{H} & \qw & \ctrl{2} & \gate{H} & \meter{} \\
        \lstick{$\ket{A}$} & \qw & \swap{1} & \swap{2} & \qw & \qw \\
        \lstick{$\ket{B}$} & \qw & \targX{} & \qw & \qw & \qw \\
        \lstick{$\ket{C}$} & \qw & \qw & \targX{} & \qw & \qw
      \end{quantikz}
  }
  \\
  \subfloat[\label{fig:multiple2}]{
      \begin{quantikz}
        \lstick{$\ket{0}$} & \gate{H} & \ctrl{3} & \qw & \gate{H} & \meter{} \\
        \lstick{$\ket{0}$} & \gate{H} & \qw & \ctrl{3} & \gate{H} & \meter{} \\
        \lstick{$\ket{A}$} & \qw & \swap{1} & \qw & \qw & \qw \\
        \lstick{$\ket{B}$} & \qw & \targX{} & \swap{1} & \qw & \qw \\
        \lstick{$\ket{C}$} & \qw & \qw & \targX{} & \qw & \qw
      \end{quantikz}
  }
  \caption{Chained SWAP tests in two possible realizations. In both cases, if the states $\ket{B}$ and $\ket{C}$ are orthogonal, the probabilities of obtaining 0 in each of the qubit measurements depends only on the inner product between the state $\ket{A}$ and the corresponding state swapped.
  }
  \label{fig:multiple}
\end{figure}
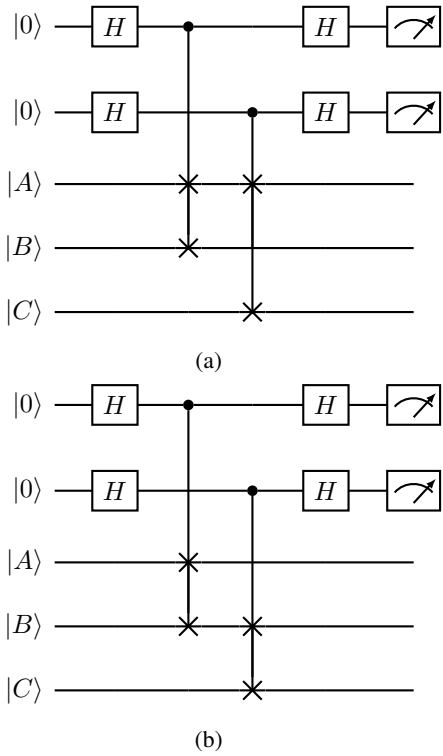

As indicated above, multiple SWAP test could be advantageous by providing the intersection of two hyperspheres as the surface for the classical search.  Fig.~\ref{fig:multiple} shows the implementation for multiple chained SWAP tests: the probability of $0$ of the top qubit is associated to the overlap $\braket{AB|BA}=|\braket{A|B}|^2$. However, the probabilities of the second bit contains (equal) contributions from $\braket{AC|CA}=|\braket{A|C}|^2$ and $\braket{BC|CB}=|\braket{B|C}|^2$. If $\ket{A}$ is the state we have prepared and $\ket{B},\ket{C}$ are the test states, when chosen to be orthogonal ($\braket{B|C}=0$), the probabilities of the second qubit will only depend on $\braket{A|C}$ (albeit with a smaller magnitude, $p(b_2=0)=\frac{1}{4}(2+|\braket{A|C}|^2+|\braket{B|C}|^2)$) so it would not be possible to exceed $p(b_2=0)=3/4$). The trade off exists between the number of state preparations and the probability above 0.5 needed to estimate the fidelity.

%% file: 03_2_Probability.tex
\section{Probability distribution}\label{sec:ProbabilityDistribution}
One important aspect of the HHL algorithm is that it belongs to that category of quantum algorithms in which the solution to the problem is not given by the most probable outcome, but instead, it is encoded within the amplitudes of the state in the basis of the measurements. This means that  ability to recreate the encoded state, and how accurately this can be done, has a direct impact on the ability to obtain the correct solution. This is a major drawback that has been stated in the original paper discussing the HHL algorithm \cite{HHL} and other papers published afterwards \cite{readthefineprint}. Building the probability distribution that contains the solution to the linear regression requires multiple ``shots" (runs) and measurements, which grows the computational complexity to reach the problem's solution.

The number of measurements necessary to extract such information varies depending on multiple factors, including the nature of the probability distribution, the smallest represented element of this distribution, and the number of potential outcomes of the measurements, which is the dimension of the problem at hand \cite{Prob1, Prob2, Prob3, Prob4}.  However, intuitively one can see how building a probability distribution of a single qubit (2 outcomes), should require fewer measurements than doing the same for, 5 qubits (32 possible outcomes) for instance. With that in mind, building the probability distribution of the ancillary bit for the controlled SWAP test should be computationally within reach, compared to the cost of recreating the actual solution state $\ket{x}$, and it is a foundational concept of the proposed approach in this paper. 

\begin{figure}
  \centering
  \subfloat[\label{fig:random}]{
    \includegraphics[width=3 in]{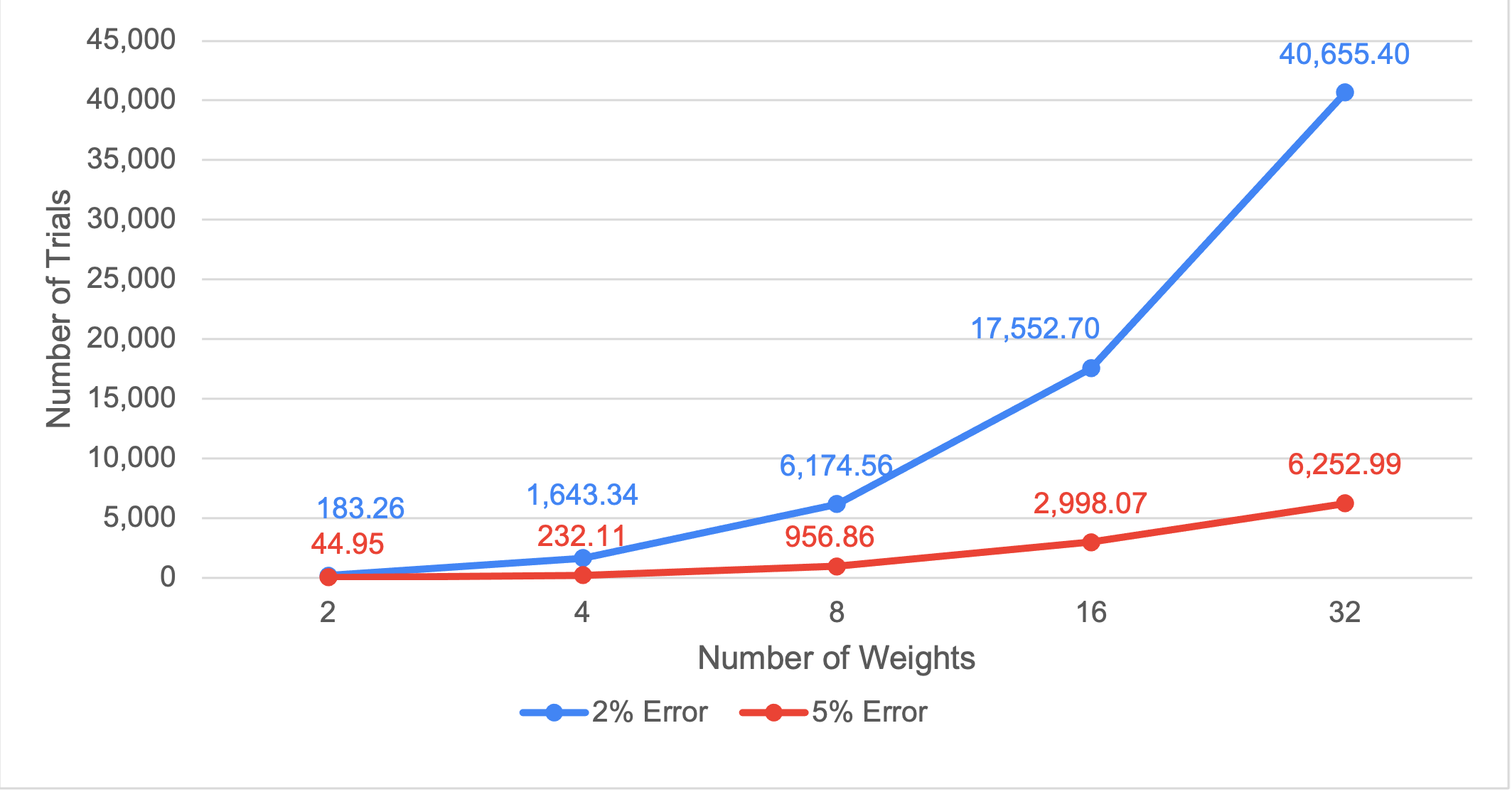}}
\\
  \subfloat[\label{fig:uniform}]{
   \includegraphics[width=3 in]{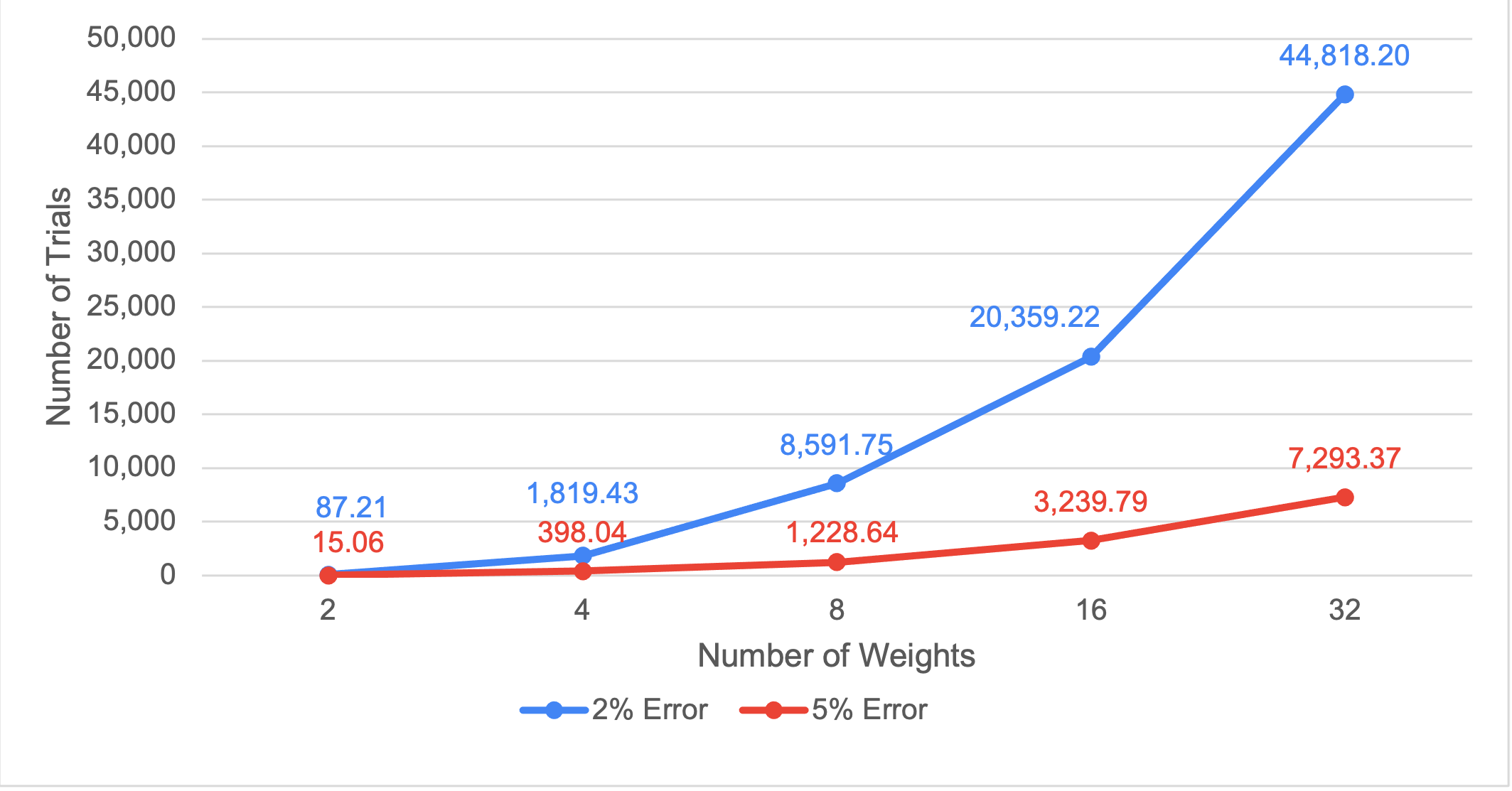}}
 \caption{Measurement number simulation to reproduce a probability distribution with L1 distance from the original probability distribution for (a) Random and (b) Uniform probability distributions. Error is meassured using the L1 distance with two possible values, 5\% and 2\%. Note that the distance between the values are not scaled, but rather just labeled. }
\end{figure}

To support the intuition above, sampling problems were simulated for two possible probability distributions, random and uniform. Multiple probability distributions of different sizes ranging from 2 to 32 were created, and samples from these were taken. The precision with which the probability distribution could be recreated was measured using the L1 distance from the original distribution. The plots of these are shown in Figures \ref{fig:random} and \ref{fig:uniform}. Each includes two plots with 2\% and 5\% L1 distance. Both show that a higher number of potential outcomes requires a significantly larger number of trials with polynomial growth. For example, for the random case, recreating the probability distribution with overall error under 2\% requires approximately 87 samples for the two outcome case, while it requires almost 45,000 samples for the 32 outcome case. Moreover, and expected, a higher precision requires a larger number of trials.

Since the weight value  for a BNN is expected to take either the values 0 or 1, reproducing the probability distribution to extract the weights can be better represented by a uniform distribution in which the binary vector has been normalized. Figure \ref{fig:uniform} can give an idea of the difficulty of recreating the values through sampling in this case. It is important to keep in mind when interpreting this figure that, although the simulations were made on fully uniform distributions ---all outcomes are equally probable---, under zero-noise condition, the number of outcomes can be interpreted as the number of 1s in the weights vector. In other words, although the data point for size 16 was taken on a uniform distribution over 16 outcomes, the same data point is also valid for, say, a 32-outcome distribution out of which 16 represent 1 values and 16 represent 0 values; or it could be the data point of a probability distribution of dimension 40 with only 16 weights equal to 1 and 24 weights equal to 0. This is due to the fact that the weights equal to zero probability are never encountered in the sampling process, as long as there is no noise.  Following this example, any probability distribution of arbitrary dimension, and with 16 equally probable events, will require over 3,000 samples to be accurately reproduced within to 5\% L1 distance overall. When noise is part of the problem, there is no guarantee that samples will not be taken out of non-zero weights, and hence, more sampling will be necessary.

In conclusion, to extract the values of the weights for a BNN using the HHL quantum algorithm, the number of measurements makes the computational cost infeasible. The SWAP test on the other hand, applied to this HHL outcome and some other wisely chosen state, requires measuring only one qubit, no matter the dimension of the swapped states. The values measured at this one single qubit containing the $\ket{d}$ state after swap being applied are not necessarily in a uniform distribution. Looking at the random distribution plot (Figure \ref{fig:random}), fewer than 90 measurements per SWAP test were shown to be enough in this experiments to get a good idea of the overlap of the two states.

%% file: 04_results.tex
\section{Experimental Results}
As a proof-of-concept, this work implemented systems of equations of size 2 and 4 successfully, simply following previously published solutions \cite{cao2012theory,pan2014experiment,bayesiandl}, with no new contribution on that end. Current Quantum Computers only allow for very small implementations of the HHL algorithm, such as those. However, the following results are presented for larger size BNN  problems, under  the assumption that the information discussed in the previous sections will be extracted quantum-mechanically for larger problem sizes in the future.

\subsection{Accelerated on MNIST Linear Regression}
\begin{figure}
 \vspace{-0.5cm}
    \centering
    \includegraphics[scale=0.5]{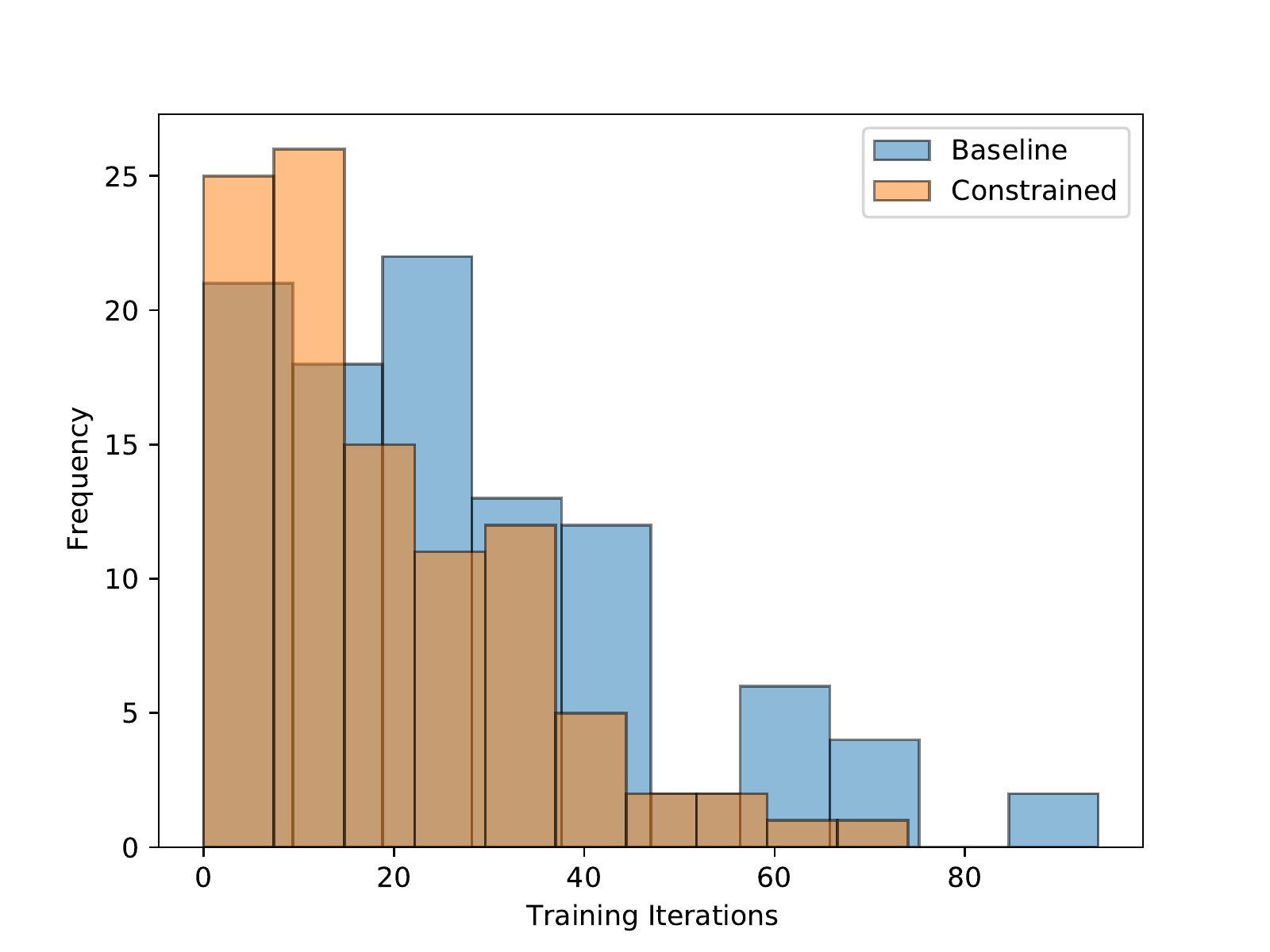}
    \caption{Number of training iterations required for convergence for unconstrained (baseline) and hyperplane constrained training of a single-layer BNN on the MNIST dataset.}
     \vspace{-0.5cm}
    \label{fig:iterations}
\end{figure}

Here, we show the results of accelerating training of a single-layer BNN for the MNIST handwritten digit dataset \cite{mnist}.  MNIST contains 60000 28$\times$28 pixel training images and 10000 test images, evenly distributed across the 10 output classes.  To match the quantum case, where linear regression is performed on a scalar-valued function, we classify each MNIST digit into only 2 classes:  `0' or not `0.'  A validation set of size 6000 was used to check for training convergence with a mean squared error loss function.  For each simulation run, the 784-inputs, 1-output BNN (with 785 weights including the bias) was first trained using the usual method of rounding each weight to the closest binary value in the forward pass and then updating the full-precision weights using gradient descent in the backward pass (see Section \ref{sec:bnn}).  When training converged, the required number of iterations and loss were recorded, as well as the solution $\mathbf{w}^{*}$.  Then, the distance $d$ from $\mathbf{w}^{*}$ to a random binary vector $\mathbf{c}$ was calculated for comparison with this new approach. This distance calculation emulates the effect of the information extracted from the HHL + SWAP test implementation explained in previous sections. A new network was generated with the same random initialization as the first, and trained using the same BNN as the original, except the weights were rounded to the closest binary point on the plane defined by $\mathbf{c}$ and $d$ (see Eq.~\eqref{eq:plane}).  After training converged, we recorded the number iterations required to achieve an equal or smaller loss value than that achieved by the original network.  This whole procedure was repeated 100 times.  The results are shown in Figure \ref{fig:iterations}.  For the original network (baseline), training converged after an average of 27.4 iterations.  For the network with weights rounded onto the hyperplane, training converged after and average of 18.6 iterations, corresponding to a $\sim$32\% improvement.  In addition, 71 out of the 100 simulations led to the constrained BNN training giving equal or better loss than the unconstrained BNN.  In the cases where the constrained BNN loss was larger, the difference in mean squared error was around 15\% of the unconstrained BNN.  While these results are encouraging, they also highlight that more work needs to be done to effectively incorporate partial information from quantum solutions to accelerate classical optimization.

%% file: 05_conclusion.tex
\section{Conclusion}
This paper proposes a form of accelerated training for Binary Neural Networks that combines quantum and classical approaches.
The quantum portion of the training applies the HHL algorithm for solving linear systems of equations to solve the linear regression of a single layer BNN. This is is followed by a SWAP test applied at the output of the HHL implementation,  to extract information out of its quantum superposed solution. Thanks to the states overlap that can be measured at the single qubit output of the SWAP test, two pieces of information can be extracted: (1) the distance with the test case, hence reducing the search to a hypersphere, and (2) the overlap with a uniform superposition, hence, the number of 1s in the weights vector.

Extracting the solution out of the HHL algorithm is computationally expensive due to the multiple measurements required. This paper shows ---through simulation of probability distributions--- that measuring the overlap represented in a single qubit after the SWAP test can be done with significantly lower number of measurements than extracting the full state, output of HHL, even when the solution state is $+\mathbb{R}$.

This information measured out of the SWAP test is used to reduce the computational complexity of the classical search for the optimal set of weights in the training process of  Binary neural networks (BNN). In the experiments on a subset of the MINST dataset, this translated into an average reduction of 32\% on the number of iterations needed for convergence. Although most of the simulations resulted in equal or better loss than the unconstrained BNN, 29\% resulted in higher loss, within 15\% MSE. These results indicate that further work should be done to explore the best way to include quantum-mechanically extracted information into the classical search.

Future continuation of this work will explore the best strategies to select the test states. In addition, although this approach was applied to BNN with 0 or 1 weights ---states that can be measured and prepared with ease--- it is possible to apply it to other ANN with weights outside this constraint. We also acknowledge that the implementation of the HHL algorithm, and quantum circuits in general,  face hurdles that have not been discussed in this paper, such as the sparsity of the matrix $A$, the approximate solution limited by the width of the QPE registers, and the multiple sources of noise. These are to be solved in a different paper.